\documentclass[pra,twocolumn,showpacs,floatfix,superscriptaddress,aps]{revtex4}
\usepackage{amsmath}
\usepackage{makeidx}
\usepackage{amssymb}
\usepackage{graphicx}

\usepackage[usenames]{color}
\usepackage[normalem]{ulem}

\usepackage{hyperref}
\usepackage[T1]{fontenc} 

\begin{document}
%%%%%%%%%%%%%%%%%%%%%%%%%%%%%%%%%%%%%%%%%%%%%%%%%%%%%%%%%%%%%%%%%%%%%%%%%%%%%%
%%%%                     Title and authors                                %%%%
%%%%%%%%%%%%%%%%%%%%%%%%%%%%%%%%%%%%%%%%%%%%%%%%%%%%%%%%%%%%%%%%%%%%%%%%%%%%%%

\title{Vector solitons in a spin-orbit coupled spin-$2$ Bose-Einstein condensate}

\author{Sandeep Gautam\footnote{sandeepgautam24@gmail.com}}
\author{S. K. Adhikari\footnote{adhikari44@yahoo.com, URL  http://www.ift.unesp.br/users/adhikari}}
\affiliation{Instituto de F\'{\i}sica Te\'orica, Universidade Estadual
             Paulista - UNESP, \\ 01.140-070 S\~ao Paulo, S\~ao Paulo, Brazil}
      
%%%%%%%%%%%%%%%%%%%%%%%%%%%%%%%%%%%%%%%%%%%%%%%%%%%%%%%%%%%%%%%%%%%%%%%%%%%%%%
%%%%%%%%%%                    Abstract                             %%%%%%%%%%%
%%%%%%%%%%%%%%%%%%%%%%%%%%%%%%%%%%%%%%%%%%%%%%%%%%%%%%%%%%%%%%%%%%%%%%%%%%%%%%

\date{\today}
\begin{abstract}
Five-component minimum-energy bound states and mobile vector solitons of a 
spin-orbit-coupled quasi-one-dimensional hyperfine-spin-2
  Bose-Einstein 
condensate are studied using the numerical solution and variational approximation of a  mean-field model. Two distinct types of solutions with single-peak and multi-peak density distribution of 
the components are identified in different domains of interaction parameters. 
From an analysis of Galilean invariance and time-reversal symmetry of the Hamiltonian,
we establish that 
vector solitons with multi-peak density distribution preserve time-reversal symmetry, but cannot propagate maintaining the shape of individual components. However, those with single-peak density 
 distribution violate time-reversal symmetry of the Hamiltonian, but can propagate 
with a constant velocity maintaining the shape of individual components.  

\end{abstract}

\maketitle

%%%%%%%%%%%%%%%%%%%%%%%%%%%%%%%%%%%%%%%%%%%%%%%%%%%%%%%%%%%%%%%%%%%%%%%%%%%%%%%
%%%%%                        Introduction                             %%%%%%%%%
%%%%%%%%%%%%%%%%%%%%%%%%%%%%%%%%%%%%%%%%%%%%%%%%%%%%%%%%%%%%%%%%%%%%%%%%%%%%%%%
\section{Introduction}
\label{Sec-I}
Bright solitons are localized wave packets that can traverse at a constant 
speed without changing their shape due to a balancing of nonlinear attraction 
and dispersive effects. Solitons have been studied in a variety of systems which 
include water waves, non-linear optics, Bose-Einstein condensates (BECs), etc 
\cite{Kivshar}. In case of BECs, the nonlinear attraction arises because of 
the net attractive inter-atomic interactions, whereas the kinetic energy term 
in the Hamiltonian is the source of dispersion. In binary mixtures of scalar 
BECs, the presence of two components allows for the emergence bright vector 
solitons \cite{Perez-Garcia}. Solitons have also been studied in spinor BECs 
with hyperfine spin $f = 1$ \cite{Ieda} and $f=2$ \cite{Uchiyama} in the absence of 
spin-orbit (SO) coupling. In a neutral atom, there is  no coupling between the 
spin and the center-of-mass motion of the atom \cite{stringari}. The SO coupling can be generated
by controlling the atom-light interaction, and hence creating artificial Abelian
and non-Abelian gauge potentials coupled to the atoms \cite{Dalibard}. 
The first experimental realization of SO coupling \cite{Lin},
with equal   Rashba \cite{Rashba} and Dresselhaus \cite{Dresselhaus} 
strengths, in a neutral atomic BEC by dressing two atomic spin states with a 
pair of lasers paved the way for other experimental studies on SO-coupled 
spinor BECs \cite{Aidelsburger}. Recently, the effect of SO coupling 
on the solitonic structures has been studied in pseudospin-$1/2$ {\cite{Achilleos, rela,Xu}} 
and spin-$1$ BECs \cite{Liu} 

 In this paper, we study an SO-coupled spin-$2$ BEC in a
quasi-one-dimensional (quasi-1D) geometry 
 characterized by three interaction strengths $
c_0\propto (4a_2+3a_4)/7$, $c_1 \propto (a_4-a_2)/7$, and 
$c_2 = (7a_0-10a_2+3a_4)/7$, where $a_0,a_2$, and $a_4$ are $s$-wave scattering 
lengths in total spin $f_{\rm tot} = 0,2$, and $4$ channels, respectively 
\cite{Koashi,Ciobanu}. Our theoretical analysis based on the SO-coupled single-particle 
Hamiltonian allows us to identify two subdomains with distinct physical properties 
separated by the $c_2=20c_1$ line. 
In the $c_2<20c_1$ subdomain,   the component densities of the wave function show a
multi-peak structure,   whereas 
in  the $c_2>20c_1$ subdomain, a single-peak structure of the same is found. Based on the 
solutions of the SO-coupled single-particle Hamiltonian, we identify an appropriate 
variational {\em ansatz} to calculate the minimum-energy bright soliton solutions for an 
SO-coupled BEC. The single-particle SO-coupled Hamiltonian is found to break 
Galilean  invariance.
By solving the Galilean transformed single-particle
Hamiltonian dynamics, we demonstrate  that  only single-peak solitons can propagate without 
changing their shape.

  In order to find the moving vector  solitons of the
SO-coupled spinor BEC, we transform the SO-coupled GP equation 
using a Galilean transformation. 
%The SO-coupled GP equation breaks Galilean invariance. 
Two degenerate solutions of the GP equation in the rest frame evolve into two velocity-dependent nondegenerate solutions in the moving frame. We show that the multi-peak soliton of Ref. \cite{Liu} is a linear combination of two such degenerate solutions in the rest frame. 
In the moving frame the degeneracy is removed, and the multi-peak soliton cannot be formed by the same linear combination. The multi-peak soliton of Ref. 
\cite{Liu} is thus velocity dependent and cannot propagate maintaining its shape. 
No such  linear combination of two degenerate solutions in the rest frame 
is needed in the formation of the single-peak    soliton identified in this paper. These
 solitons 
{ can} emerge as the ground states of the coupled GP 
equation in a moving frame, and hence  
can move with a constant velocity maintaining their shape, including the densities and phase profiles of individual components, 
unlike the moving multi-peak solitons of Refs. \cite{rela,Liu} which show spin-mixing 
dynamics. %{ Single-peak bright solitons can also move without
%changing their shape in quasi-1D pseudospin-$1/2$ BECs with Rashba
%type SO coupling \cite{Sakaguchi}.} 

 In Sec. \ref{Sec-II}, we describe the mean-field
formalism used to study the SO-coupled spin-$2$ BECs. Here, taking into
account the effect of interactions on the solutions of the non-interacting and
trapless system, we identify two types of ground state solutions with single-peak and multi-peak 
densities, respectively. In case of ground-state solutions with 
single-peak densities, the reduction of the coupled GP equation to a single nonlinear
partial differential equation using single-mode approximation (SMA) is also discussed in this section. 
In SMA all the component wavefunctions are assumed to have the same spatial dependence but 
can have different
normalization \cite{Law}. In 
Sec. \ref{Sec-III}, we study the minimum-energy stationary %and moving bright soliton 
%solution of the 
solitons by variationally minimizing the energy. We also investigate the 
moving solitons from a consideration of Galilean invariance of the Hamiltonian. 
In Sec. \ref{Sec-IV}, we present the numerical results for 
both types of solitons.
We conclude by providing a summary of the study in Sec. \ref{Sec-V}.
%%%%%%%%%%%%%%%%%%%%%%%%%%%%%%%%%%%%%%%%%%%%%%%%%%%%%%%%%%%%%%%%%%%%%%%%%%%%%%%
%%%%%            Mean-field model for a SO-coupled BEC                %%%%%%%%%
%%%%%%%%%%%%%%%%%%%%%%%%%%%%%%%%%%%%%%%%%%%%%%%%%%%%%%%%%%%%%%%%%%%%%%%%%%%%%%%
\section{Spin-orbit coupled BEC in a  quasi-1D trap}
\label{Sec-II}  

Although, unlike in the case of charged electrons in an atom, an SO 
coupling interaction does not naturally arise in the case of neutral atoms in a 
spinor BEC, an SO coupling  can be realized   by an 
engineering of atom-light interaction. 
An SO-coupled spinor BEC in a quasi-1D trap can be realized by making 
the trapping frequencies along the two axes, say $y$ and $z$ ($\omega_y,\omega_z$), much larger than 
that along the $x$ axis ($\omega_x$). This leads to a BEC in which the dynamics
is restricted only along $x$ axis. The single-particle 
Hamiltonian of this BEC with   in a quasi-1D trap is \cite{Lin,zhai}
\begin{equation}
H_0 = \frac{p_x^2}{2m} + V(x) + \gamma p_x \Sigma_x,
\label{sph} 
\end{equation}
where $p_x = -i\hbar\partial/\partial x$ is the momentum operator along x
axis, $V(x)=m\omega_x^2x^2/2$ is the harmonic trapping potential along $x$ axis, and
$\Sigma_x$ is the irreducible representation of $x$ component of the  spin operator:
\begin{eqnarray}
\Sigma_x= \left( \begin{array}
 {ccccc}
0 & 1 & 0 & 0 & 0\\
1 & 0 & \sqrt{3/2} & 0 & 0\\
0 & \sqrt{3/2} & 0 & \sqrt{3/2} & 0\\
0 & 0 & \sqrt{3/2} & 0 & 1\\
0 & 0 & 0 & 1 & 0\end{array} \right) .
\end{eqnarray}
The SO coupling corresponds to equal strengths of Rashba 
($\Sigma_x p_x+\Sigma_y p_y)$  and Dresselhaus ($\Sigma_x p_x - \Sigma_y p_y$)  
couplings \cite{H_Zhai,Kawaguchi,zhai}, where $\Sigma_y$ is the irreducible representation of $y$ 
component of the spin operator.   An equivalence of these couplings with  the usual  Rashba 
($\Sigma_x p_y-\Sigma_y p_x)$ \cite{Rashba} and Dresselhaus ($\Sigma_x p_y + \Sigma_y p_x$)  
\cite{Dresselhaus}
couplings can be shown by a rotation in the spin space \cite{H_Zhai,zhai}.

In our previous studies \cite{gautam-1,gautam-2},
we employed a different SO coupling ($p_x\Sigma_z$) again with equal 
  Rashba \cite{Rashba} and Dresselhaus \cite{Dresselhaus} strengths. 
The present 
single-particle Hamiltonian (\ref{sph}) 
already couples the different spin components. The previous SO interaction 
\cite{gautam-1,gautam-2} 
did not couple the different spin components at the single-particle level and the 
coupling was achieved after introducing the nonlinear spinor interactions in the BEC.

%We consider an SO-coupled spinor condensate in a quasi-1D trap in which the
%trapping frequencies along the $y$ and $z$ axes are much larger than that 
%along the $x$ axis \cite{Salasnich}. The single particle Hamiltonian of the 
%condensate with equal strengths of Rashba \cite{Rashba} and Dresselhaus 
%\cite{Dresselhaus} SO couplings in such a quasi-1D trap is \cite{H_Zhai} 
%\begin{equation}
%H_0 = \frac{p_x^2}{2m} + V(x) + \gamma p_x \Sigma_x,
%\label{sph} 
%\end{equation}
%where $p_x = -i\hbar\partial/\partial x$ is the momentum operator along x
%axis, $V(x)=m\omega_x^2x^2/2$ is the harmonic trapping potential along $x$ 
%axis, 
%and $\Sigma_x$ is the irreducible representation of $x$ component of 
%spin-$2$ matrix

Taking interactions in the Hartree approximation and using the single-particle 
Hamiltonian  (\ref{sph}), a quasi-1D spin-$2$ BEC 
can be described by the following set of five coupled mean-field partial 
differential equations for the wave-function components $\psi_j$ 
\cite{Kawaguchi}
\begin{align}
 i\hbar&\frac{\partial \psi_{\pm 2}}{\partial t} =
 \left( -\frac{\hbar^2}{2m}\frac{\partial^2}{\partial x^2}
 +V(x)+c_0\rho\right)\psi_{\pm 2}-i\hbar\gamma
  \frac{\partial\psi_{\pm 1}}{\partial x}\nonumber\\ &
 +c_1\big(F_{\mp} \psi_{\pm 1} \pm 2F_z\psi_{\pm 2}\big) 
  +\big({c_2}/{\sqrt{5}}\big)\Theta\psi_{\mp 2}^*, \label{gp-1}\\
i\hbar&\frac{\partial \psi_{\pm 1}}{\partial t} =
 \left( -\frac{\hbar^2}{2m}\frac{\partial^2}{\partial x^2}
 +V(x)+c_0\rho\right)\psi_{\pm 1}-i\hbar\gamma
  \frac{\partial\psi_{\pm 2}}{\partial x}  \nonumber\\
  & -i\hbar\gamma\frac{\sqrt{6}}{2}\frac{\partial\psi_{0}}{\partial x} +
  c_1\big(\sqrt{3/2}F_{\mp}\psi_0+F_{\pm}\psi_{\pm 2}\pm F_z\psi_{\pm 1}\big)
  \nonumber\\
  & -\big({c_2}/\sqrt{5}\big)\Theta\psi_{\mp 1}^*\label{gp-2},\\
i\hbar &\frac{\partial \psi_0}{\partial t} = 
 \left( -\frac{\hbar^2}{2m}\frac{\partial^2}{\partial x^2}
 +V(x)+c_0\rho\right)\psi_0 -i\hbar\gamma\frac{\sqrt{6}}{2}
 \Bigg(\frac{\partial\psi_{1}}{\partial x}  \nonumber\\
  &+\frac{\partial\psi_{-1}}{\partial x}\Bigg)+
 \frac{\sqrt{6}}{2}c_1\big(F_{-} \psi_{-1}+F_{+}\psi_1\big) 
  + \frac{c_2}{\sqrt{5}}\Theta\psi_{0}^*,\label{gp-3}
\end{align}
where $c_0 = 2\hbar^2 (4a_2+3a_4)/(7m l_{yz}^2)$, 
$c_1 = 2\hbar^2 (a_4-a_2)/(7m l_{yz}^2)$, 
$c_2 = 2\hbar^2 (7a_0-10a_2+3a_4)/(7m l_{yz}^2)$,  
$a_0$, $a_2$ and $a_4$ are the $s$-wave scattering lengths in the total 
spin $f_{\mathrm{tot}} = 0,2$ and $4$ channels, respectively,    
$\rho_j(x) = |\psi_j|^2$ with $j=2,1,0,-1,-2$ are the component densities, 
$\rho(x)= \sum_{j=-2}^2\rho_j$ is the total density, 
and $l_{yz} = \sqrt{\hbar/(m\omega_{yz})}$ with  
$ \omega_{yz}=\sqrt{\omega_y\omega_z}$ is the oscillator length in the 
transverse $y-z$ plane and
\begin{align}
F_{+} = F_{-}^*&=F_x+i F_y\nonumber\\
               &=2(\psi_2^*\psi_1+\psi_{-1}^*\psi_{-2})% \nonumber \\ &
+\sqrt{6}(\psi_1^*\psi_0
+\psi_0^*\psi_{-1}),\nonumber \\
F_z &= 2(|\psi_2|^2-|\psi_{-2}|^2) + |\psi_1|^2-|\psi_{-1}|^2,\nonumber \\
 \Theta &= \frac{2\psi_2\psi_{-2}-2\psi_1\psi_{-1}+\psi_0^2}{\sqrt{5}},\nonumber 
\end{align}
where $F_x,F_y$, $F_z$ are the three components of the spin-density
vector $\mathbf F$, and $\Theta$ is the spin-singlet pair amplitude \cite{Kawaguchi}. 
Before proceeding further, let us transform the 
Eqs. (\ref{gp-1})-(\ref{gp-3}) into dimensionless form using 
\begin{equation}
 \tilde{t} = \omega_x t,~\tilde{x} = \frac{x}{l_0},
 ~\phi_j(\tilde{x},\tilde{t}) = 
 \frac{\sqrt{l_0}}{\sqrt{N}}\psi_j(\tilde{x},\tilde{t}), 
\end{equation}
where $l_0=\sqrt{\hbar/(m\omega_{x}}$) is the oscillator length along $x$ 
axis, and $N$ is the total number of atoms. Then,   Eqs. 
(\ref{gp-1})-(\ref{gp-3}) in dimensionless form become
\begin{align}
 i &\frac{\partial \phi_{\pm 2}}{\partial \tilde{t}} =
 \left( -\frac{1}{2}\frac{\partial^2}{\partial \tilde{x}^2}
 +\tilde{V}(\tilde{x})+\tilde{c}_0\tilde{\rho}\right)\phi_{\pm 2}-i\tilde{\gamma}
  \frac{\partial\phi_{\pm 1}}{\partial \tilde{x}}\nonumber\\ &
 +\tilde{c}_1\big(\tilde{F}_{\mp} \phi_{\pm 1} 
 \pm 2\tilde{F}_{\tilde{z}}\phi_{\pm 2}\big) 
  +\big({\tilde{c}_2}/{\sqrt{5}}\big)\tilde{\Theta}\phi_{\mp 2}^*, \label{gps-1}\\
i&\frac{\partial \phi_{\pm 1}}{\partial \tilde{t}} =
 \left( -\frac{1}{2}\frac{\partial^2}{\partial \tilde{x}^2}
 +\tilde{V}(\tilde{x})+\tilde{c}_0\tilde{\rho}\right)\phi_{\pm 1}-i\tilde{\gamma}
  \frac{\partial\phi_{\pm 2}}{\partial \tilde{x}}  \nonumber\\
  & -i\tilde{\gamma}\frac{\sqrt{6}}{2}\frac{\partial\phi_{0}}{\partial \tilde{x}} +
  \tilde{c}_1\big(\sqrt{3/2}\tilde{F}_{\mp}\phi_0+
 \tilde{F}_{\pm}\phi_{\pm 2}\pm \tilde{F}_{\tilde{z}}\phi_{\pm 1}\big)\nonumber\\
  & -\big({\tilde{c}_2}/\sqrt{5}\big)\tilde{\Theta}\phi_{\mp 1}^*\label{gps-2},
\end{align}
\begin{align}
i &\frac{\partial \phi_0}{\partial \tilde{t}} = 
 \left( -\frac{1}{2}\frac{\partial^2}{\partial \tilde{x}^2}
 +\tilde{V}(\tilde{x})+\tilde{c}_0\tilde{\rho}\right)\phi_0 -i\tilde{\gamma}
 \frac{\sqrt{6}}{2}\Bigg(\frac{\partial\phi_{1}}{\partial \tilde{x}}  \nonumber\\
  &+\frac{\partial\phi_{-1}}{\partial \tilde{x}}\Bigg)+
\frac{\sqrt{6}}{2}\tilde{c}_1\big(\tilde{F}_{-} \phi_{-1}+\tilde{F}_{+}\phi_1\big) 
  + \frac{\tilde{c}_2}{\sqrt{5}}\tilde{\Theta}\phi_{0}^*,\label{gps-3},
\end{align}
where $\tilde{V} = \tilde{x}^2/2$, 
$\tilde{\gamma} = \hbar k_r/(m\omega_x l_0)$, 
$\tilde{c}_0 = 2N (4a_2+3a_4)l_0/(7l^2 _{yz})$,
$\tilde{c}_1 = 2N (a_4-a_2)l_0/(7l^2 _{yz})$, 
$\tilde{c}_2 = 2N (7a_0-10a_2+3a_4)l_0/(7l^2_{yz})$,
$\tilde{\rho}_j(\tilde{x}) = |\phi_j|^2$ with $j=2,1,0,-1,-2$, 
	and $\tilde{\rho}(\tilde{x}) = \sum_{j=-2}^2|\phi_j|^2$
and
\begin{align}
\tilde{F}_{+} = \tilde{F}_{-}^*=& 2(\phi_2^*\phi_1+\phi_{-1}^*\phi_{-2})%\nonumber\\& 
+\sqrt{6}(\phi_1^*\phi_0 +\phi_0^*\phi_{-1}),  \\
\tilde{F}_{\tilde{z}} =& 2(|\phi_2|^2-|\phi_{-2}|^2) + |\phi_1|^2-|\phi_{-1}|^2,  \\
\tilde{\Theta} =& \frac{2\phi_2\phi_{-2}-2\phi_1\phi_{-1}+\phi_0^2}{\sqrt{5}}. 
\end{align}
The total density is now normalized to unity, i.e., 
$
 \int_{-\infty}^{\infty} \tilde{\rho}(\tilde{x})d\tilde{x} = 1.
$
We will represent the dimensionless variables without tildes in the rest of
the manuscript for notational simplicity.

\begin{figure}
    \includegraphics[trim = 5mm 0mm 0cm 0mm, clip,width=0.8\linewidth,clip]{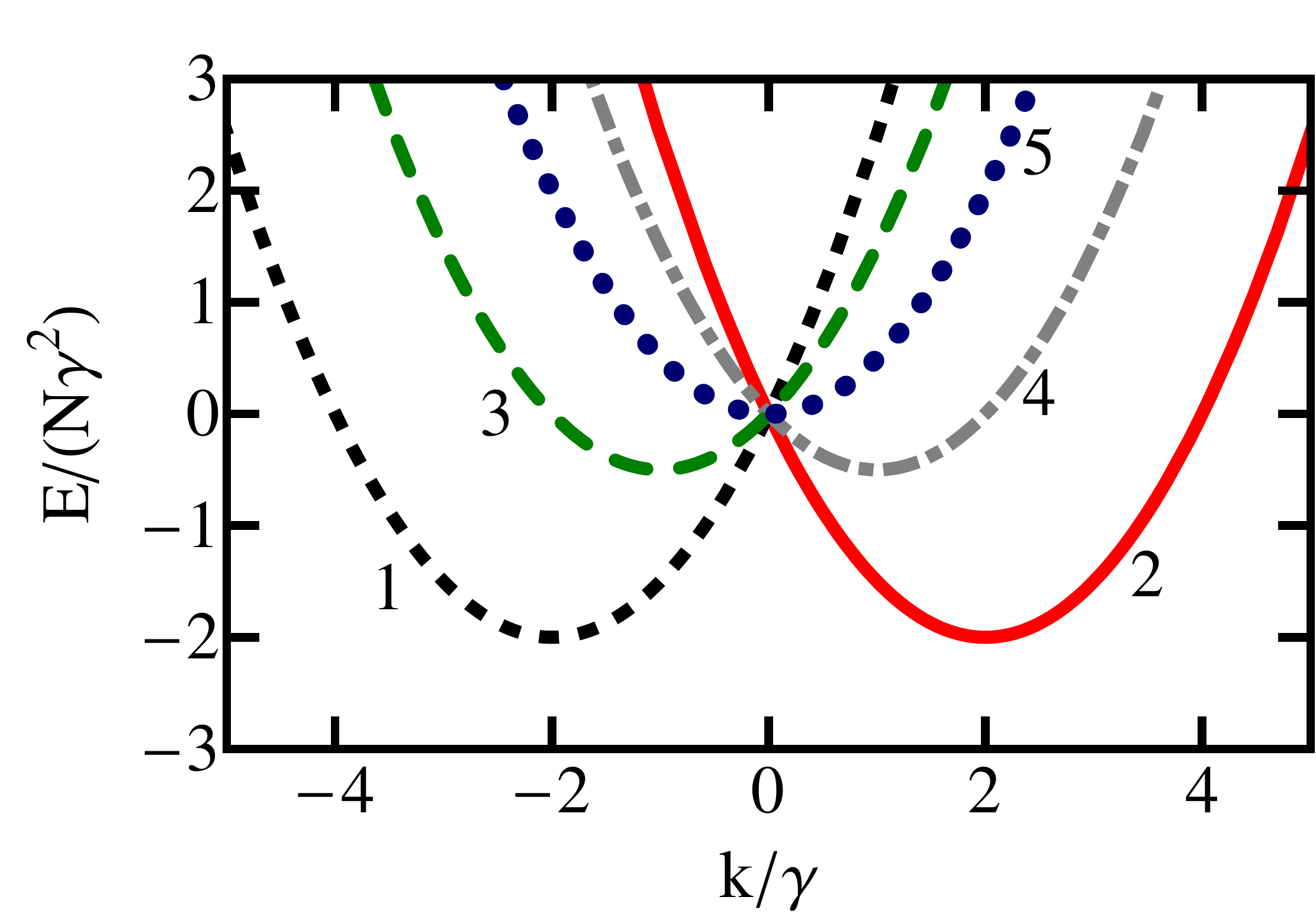}
\caption{(Color online) Energy dispersion $E/(N\gamma^2)$ vs.  $k/\gamma$  for the eigen 
functions of the single-particle SO-coupled Hamiltonian for the five wave 
functions $\Phi_i$, where  $i=1,...,5.$ }
 \label{fig1}
\end{figure}

 For a non-interacting system in
the absence of a trapping potential, there are five linearly independent solutions
of Eqs. (\ref{gps-1})-(\ref{gps-3}): 
%$E_{\rm min} = -2N\gamma^2$ 
%two linearly independent solutions
%of Eqs. (\ref{gps-1})-(\ref{gps-3}) corresponding to a minimum total energy 
%$E_{\rm min} = -2N\gamma^2$
\begin{eqnarray}
\Phi_1&=&  \frac{e^{i k x}}{4}\left( 
1,2,\sqrt{6},2,1 \right)^T, 
\label{eigen_function1}\\ 
\Phi_2&=& \frac{e^{i k x}}{4}\left( 
 1,-2,\sqrt{6},-2,1  \right)^T,
\label{eigen_function2} \\
\Phi_3&=& \frac{e^{i k x}}{2}\left( 
 -1,-1,0,1,1  \right)^T,
\label{eigen_function3} \\
\Phi_4&=& \frac{e^{i k x}}{2}\left( 
 -1,1,0,-1,1  \right)^T,
\label{eigen_function4} \\
\Phi_5&=&  e^{ i k x} \sqrt{\frac{3}{8}}\left( 
 1,0,-\sqrt{\frac{2}{3}},0,1  \right)^T,
\label{eigen_function5}
\end{eqnarray}
where $T$ stands for transpose. 
Here $\Phi_1$ and $\Phi_2$ are two degenerate 
solutions corresponding to energy per particle $E/N=k^2/2\pm 2k\gamma,$ which has a minimum  $E_{\rm min}/N = -2\gamma^2$ 
at $k=\mp 2 \gamma${, where the upper and lower signs are for $\Phi_1$ and $\Phi_2$, respectively}.
Similarly, the eigen functions $\Phi_3$ and $\Phi_4$ are also degenerate with energy 
$E/N = k^2/2\pm k\gamma, $ which has a minimum  $E_{\rm min}/N = -\gamma^2/2$ at $k=\mp \gamma$ { with
upper and lower signs for $\Phi_3$ and $\Phi_4$, respectively;} 
whereas the non-degenerate eigen function $\Phi_5$ 
has energy $E/N = k^2/2, $ which has a minimum $E_{\rm min}/N = 0$ at $k=0$.  The $E/(N\gamma^2)$ vs. $k/\gamma$ 
dispersion curves are shown in Fig. \ref{fig1} and we will consider the lowest-energy ground state 
$\Phi_{1, 2}$ with $k=\mp 2\gamma.$

Compared to the SO-coupled pseudospin-$1/2$ \cite{C_Wang,Ho} { and spin-1 condensates \cite{C_Wang}}, there 
are two sets of degenerate eigen functions in the spin-2 case. Due to this, two distinct types of
muti-peak stationary profiles can 
emerge in SO coupled spin-2 condensate.

 The most general
solution of Eqs. (\ref{gps-1})-(\ref{gps-3}) with minimum energy 
for a BEC with a uniform density $n$ in the absence of interactions and trapping 
potential can be obtained by the 
linear superposition of $\sqrt{n}\Phi_1$
and $\sqrt{n}\Phi_2$
\begin{eqnarray}\label{abc}
\sqrt {n} \Phi &=& \left( \begin{array}{c}
\phi_2 \\
\phi_{1}\\
\phi_{0}\\
\phi_{-1}\\
\phi_{-2}\\ \end{array} \right) = \sqrt{n}\left(\alpha_1 \Phi_1 + \alpha_2 \Phi_2\right),\\
    &=& \frac{\sqrt{n}}{4}\left( \begin{array}{c}
\alpha_1e^{-i2\gamma x} + \alpha_2e^{i2\gamma x}\\
2\alpha_1e^{-i2\gamma x} -2 \alpha_2e^{i2\gamma x}\\
\sqrt{6}\alpha_1e^{-i2\gamma x} +\sqrt{6} \alpha_2e^{i2\gamma x}\\
2\alpha_1e^{-i2\gamma x} -2 \alpha_2e^{i2\gamma x}\\
\alpha_1e^{-i2\gamma x} + \alpha_2e^{i2\gamma x} \end{array}\right),
\label{single_part_sol}
\end{eqnarray}
where $|\alpha_1|^2+|\alpha_2|^2=1$, so that $\Phi$ is normalized to unity. 
The solution  (\ref{single_part_sol}) implies that the 
magnetization ${\cal M}\equiv  \int F_z dz = 0$. It implies
that we can only obtain stable bright solitons with zero magnetization for
the SO-coupled spinor condensate considered in this paper. 
On the other hand, in the absence of SO coupling, one can have stable 
bright solitons with non-zero magnetization for spin-2 condensate \cite{Uchiyama}.

The time-reversal-symmetric single-particle Hamiltonian (\ref{sph}) violates  
parity.  This will have interesting consequences on its eigen functions. 
The effect of time-reversal symmetry operator ${\cal T}$ acting on a spin or orbital  
angular-momentum state $|j,m\rangle$ is \cite{Kawaguchi} 
\begin{equation}
({\cal T} \phi)_m=(-1)^m \phi_m^*.
\end{equation}
  Time-reversal symmetry of a quantum state $\Phi$ requires
that it should be the same as its time-reversed state apart from a phase factor. 
The degenerate states (\ref{eigen_function1}) and (\ref{eigen_function2}) violate time-reversal symmetry and 
are connected by time-reversal symmetry operator:  
$ {\cal T}  |\Phi_1\rangle = - |\Phi_2\rangle, {\cal T}  |\Phi_2\rangle = - |\Phi_1\rangle. $ In general, for arbitrary $\alpha_1$ and $\alpha_2$ the states (\ref{single_part_sol}) do not have  
time-reversal symmetry.

In the presence of interactions, the interaction energy per 
particle is \cite{Kawaguchi}
\begin{eqnarray}
\epsilon_{\rm int} &=& \frac{c_0}{2}n + \frac{c_1}{2n}|\mathbf F|^2 + 
                       \frac{c_2}{2n}|\Theta|^2,\nonumber\\
                   &=& \frac{c_0}{2}n +2n\left[c_1
+|\alpha_1|^2|\alpha_2|^2\left(\frac{c_2-20c_1}{5}\right)\right]. 
\end{eqnarray}
If $c_2<20c_1$, the minimum of $\epsilon_{\rm int}$ 
corresponds to $|\alpha_1| = |\alpha_2| = 1/\sqrt{2}$. This state is nondegenerate and has 
multi-peak density distribution for the wave-function components
  and is time-reversal
invariant. The spin expectation and absolute value of spin-singlet pair amplitude 
per particle for this state are, respectively, $|\mathbf F|/n = 0$ and $|\Theta|/n =1/\sqrt{5}$.
On the other hand, for $c_2>20c_1$, the $\epsilon_{\rm int}$ can be minimized if $|\alpha_1| = 1, 
|\alpha_2| = 0$ or $|\alpha_1| = 0, |\alpha_2| = 1$. These two plane-wave states are degenerate 
and are connected by the time-reversal operator  and hence    
 break time-reversal symmetry of the Hamiltonian.  
The spin expectation per particle for this state $|\mathbf F|/n = 2$,
and the absolute value of spin-singlet pair amplitude per particle $|\Theta|/n = 0$.
%This state is also called ferromagnetic phase \cite{Koashi,Ciobanu,Kawaguchi} and
%leads to the 
These states have single-peak density distribution for the wave-function components
and can also be studied using SMA. In SMA, the order parameter can be
written as
\begin{equation}
\Phi(x,t) = \frac{1}{4}  (1,2,\sqrt{6},2,1)^T \phi_{\rm SMA}(x,t),
\label{sma}
\end{equation} % which assumes that the different
%spin states can be represented by the same wavefunction $\phi_{\rm SMA}$. 
%Here $\zeta$ is the normalized spinor and is given by
%\begin{equation}
%\zeta = (1,2,\sqrt{6},2,1)/4
%\label{zeta_spinor}
%\end{equation}
based on the minimum-energy solutions (\ref{eigen_function1})  
of the single-particle Hamiltonian. For SMA to be valid, all the spin-dependent interactions should be much
smaller than the spin-independent interactions \cite{Kawaguchi}. Hence, using Eq. (\ref{sma}) % and (\ref{zeta_spinor}) 
in Eqs. (\ref{gps-1})-(\ref{gps-3}) and neglecting the $c_1$- and $c_2$-dependent terms, one can obtain
the single nonlinear differential equation
\begin{eqnarray}
 i \frac{\partial \phi_{\rm SMA}}{\partial {t}} =
 \left[ -\frac{1}{2}\frac{\partial^2}{\partial {x}^2}
 +{V}({x})+{c}_0{\rho  -2i{\gamma}
  \frac{\partial }{\partial {x}}}\right]\phi_{\rm SMA}%\nonumber\\
\label{sma_eq},
\end{eqnarray}
where the tildes have been dropped. Equation (\ref{sma_eq}) will be solved numerically.
%Besides the $\gamma$-dependent term, this equation is nothing but the GP equation for 
%a scalar BEC and can be solved numerically \cite{Muruganandam}.

\section{Bright solitons}
\label{Sec-III}

\subsection{Variational analysis}
\label{Sec-IIIA}
We just found that the physical properties of the system are different for $c_2>20c_1$ and 
$c_2<20c_1$, and now we present a variational analysis of the bright soliton with minimum energy 
in these two domains.
  We variationally minimize the energy 
of the trapless BEC \cite{Kawaguchi,Kawakami}, given by
\begin{align}
 E &=  N\int_{-\infty}^{\infty} \Bigg\{\frac{1}{2}\sum_{j=-2}^2 \left|\frac{d\phi_j}{dx}\right|^2
  -i\gamma \phi_2^*\frac{d\phi_1}{dx} -i\gamma \phi_{-2}^*\frac{d\phi_{-1}}{dx}\nonumber\\
  &-i\gamma\frac{\sqrt{6}}{2}\phi_0^*\left(\frac{d\phi_1}{dx} + \frac{d\phi_{-1}}{dx}\right)
  -i\gamma \phi_1^*\frac{d\phi_2}{dx} -i\gamma \phi_{-1}^*\frac{d\phi_{-2}}{dx}\nonumber\\
  &-i\gamma\frac{\sqrt{6}}{2} \frac{d\phi_0}{dx}\left(\phi_1^*+\phi_{-1}^*\right)\nonumber\\
  &+  \frac{{c_0}\rho^2 + {c_1}|\mathbf F|^2+{c_2}|\Theta|^2}{2}
 \Bigg\}dx
  \label{energy},
\end{align}
in the two domains which are separated by the $c_2=20c_1$ line. An appropriate 
variational {\em ansatz} $\Phi_{\rm var}$ is constructed by taking the product of the
 superposition  of the eigen functions corresponding to minimum energy
 in Eqs. (\ref{eigen_function1}) and  (\ref{eigen_function2}) 
with a localized spatial
soliton, i.e.,
{
\begin{equation}
\Phi_{\rm var} = \frac{\sqrt{\sigma}}{2}\Phi \rm {sech} 
                 (\sigma x),
\label{var_ansatz}  
\end{equation}
}
where $\sigma$ is the variational parameter characterizing the width and strength
of the soliton, and $|\alpha_1|^2+|\alpha_2|^2 = 1$. As discussed in Sec. \ref{Sec-II},
the exact values of $\alpha_1$ and $\alpha_2$ differ in the two domains.
In order to search for the stable stationary bright solitons with 
higher energies, one can also consider the following variational {\em ansatz}:
\begin{eqnarray}
\Phi_{\rm var} &=& \frac{\sqrt{\sigma}}{2}(\alpha_1\Phi_3 + \alpha_2\Phi_4) \rm {sech} 
                 (\sigma x),%~~ {\rm and}
\label{var_ansatz2}\\
\Phi_{\rm var} &=& \frac{\sqrt{\sigma}}{2}\Phi_5 \rm {sech} 
                 (\sigma x)
\label{var_ansatz3}.
\end{eqnarray}
Ansatz (\ref{var_ansatz2}) and (\ref{var_ansatz3}) can lead to   bright solitons
with successively higher energies than the one described by Eq. (\ref{var_ansatz}).
Nevertheless, in the present paper, our emphasis is to calculate the minimum-energy bright 
solitons consistent with Eq. (\ref{var_ansatz}).

If {\em $c_2<20c_1$}, we choose $|\alpha_1| = |\alpha_2| = 1/\sqrt{2}$
in Eq. (\ref{var_ansatz}). Then, the energy of the soliton is
\begin{equation}
E = \frac{N}{30} \left(-60 \gamma ^2+5 \sigma ^2+5 \sigma  c_0+\sigma  c_2\right)
\end{equation} 
with a minima  at
\begin{equation}
\sigma  = -\frac{1}{10} \left(5 c_0+c_2\right),
\label{sw1}
\end{equation}
provided that $5 c_0+c_2<0$.  The energy and hence the shape of the
soliton are not the functions of interaction parameter $c_1$ due the fact
$\mathbf F = 0$ for $|\alpha_1| = |\alpha_2| = 1\sqrt{2}$ as has been
discussed in Sec. \ref{Sec-II}.  This parameter domain corresponds to the 
antiferromagnetic phase ($\mathbf F=0,  |\Theta|/n =1/\sqrt{5}$) \cite{Ciobanu}. 
{
The state $\Phi_{\rm var}$, defined by Eqs. (\ref{var_ansatz}) and (\ref{single_part_sol})
 with $|\alpha_1| = |\alpha_2| = 1/\sqrt{2}$ is time-reversal symmetric.}

If {\em $c_2>20c_1$}, we choose $|\alpha_1| = 1$, $|\alpha_2| = 0$ or vice versa
in Eq. (\ref{var_ansatz}). The energy of the soliton, then, is 
\begin{equation}
E = \frac{N}{6} \left(-12 \gamma ^2+\sigma ^2+\sigma  c_0+4 \sigma  c_1\right)
\end{equation}
with a minima at
\begin{equation}
\sigma = -\frac{1}{2} \left(c_0+4 c_1\right),
\label{sw2}
\end{equation}
provided  that $c_0+4 c_1 < 0$.  The shape of the
soliton is independent of interaction parameter $c_2$ due vanishing
$\Theta$. This parameter domain corresponds to the 
ferromagnetic phase ($|\mathbf F|/n=2,  \Theta =0)$ \cite{Ciobanu}. 
{
The $\Phi_{\rm var}$ with $|\alpha_1| = 1$, $|\alpha_2| = 0$ 
or vice versa leads to single-peak density distribution for wave-function
components and breaks time-reversal symmetry. 

\subsection{Moving bright solitons}
\label{III-B}
The GP equation governing the dynamics and statics of a scalar BEC is Galilean
invariant. The implication of this invariance can be understood by considering the 
scalar 1D GP equation in dimensionless form
\begin{equation}\label{gp}
i \frac{\partial \phi(x,t)}{\partial t}= -\frac{1}{2}\frac{\partial^2 \phi(x,t)}{\partial x^2}-c|\phi(x,t)|^2  \phi(x,t),
\end{equation}
where the nonlinearity is considered attractive ($c>0$) for the formation of a bright soliton. This equation has the analytic solution
\begin{equation}\label{mvsol}
\phi(x,t)=\sqrt{({\sigma/ c})} \mathrm{sech}\left[(x-vt)\sqrt\sigma \right] e^{ivx+i(\sigma-v^2)t/2},
\end{equation}
where $v$ is the velocity and $\sigma$ represents the width and the strength of the soliton. It implies that the stationary soliton   $\phi(x,0)= \sqrt{({\sigma/ c})} \mathrm{sech}\left[x\sqrt\sigma \right] $ moves as the  bright soliton 
 $\phi_M(x,t)$ defined by \cite{rela}
\begin{equation}\label{mulxy}
\phi_M(x,t)= \phi(x-vt,t) e^{ivx+i(\sigma-v^2)t/2},
\end{equation}
maintaining the width and strength  ($\sigma$) fixed.  
In the case of a multi-component spinor BEC, Eq. (\ref{mulxy}) remains valid with 
a multi-component wave function $\Phi$ replacing the single-component $\phi$ while 
the multiplying exponential factor remains unchanged.  

%In the presence of SO coupling, the Galelian invariance is no longer ensured since the 
%SO-coupled Hamiltonian is no longer Galilean invariant as in the case of a  
%pseudospin-$1/2$ \cite{rela,Sakaguchi} and spin-$1$ BEC \cite{gautam-3}.
Now, let us examine the Galilean invariance of the SO-coupled Hamiltonian. 
Equation (\ref{mulxy}) implies that for the Galilean transformation $x'=x+vt, t'=t$,
where $v$ is the relative velocity of unprimed coordinate system with respect
to primed coordinate system, 
the wave function $\Phi$ of Eqs. (\ref{gps-1})-(\ref{gps-3}) should transform to 
$\Phi_M$ related by    
\begin{equation}\label{mulxyz}
\Phi(x,t)= \Phi_M(x',t') e^{-ivx'-i(\sigma-v^2)t'/2}.
\end{equation}
Substituting Eq.  (\ref{mulxyz})  
in Eqs. (\ref{gps-1})-(\ref{gps-3}) and using
$ \partial/\partial x= \partial/\partial x', \partial/\partial t =  \partial/\partial t' + v\partial/\partial x'$, we get
\begin{equation}
i\frac{\partial \Phi_M(x',t')}{\partial t'} = \left[-\frac{1}{2}\frac{\partial^2}{\partial x'^2}-
\gamma\Sigma_x\left(i\frac{\partial}{\partial x'}+v\right)\right]\Phi_M(x',t'),
\label{mov_sol}
\end{equation}
where, for the sake of simplicity, we have suppressed the terms proportional to 
$c_0 $, $c_1$, and $c_2$, which should remain unchanged. We have also dropped a $\sigma$ 
dependent additive term in the Hamiltonian which does not contribute to the dynamics.

The extra term 
$-\gamma\Sigma_x v\Phi_M$ on the right hand side of the equation indicates that 
the SO-coupled Hamiltonian is no longer Galilean invariant. 
The  SO-coupled equation (\ref{mov_sol}) has the solutions $\Phi_1$ and $\Phi_2$ 
of Eqs. (\ref{eigen_function1}) and (\ref{eigen_function2}) with energies  $E=-2N(v \gamma + \gamma ^2)$
and $E=2N(v \gamma - \gamma ^2)$, respectively. Hence,   the degeneracy between $\Phi_1$ and $\Phi_2$ is 
removed for $v\ne 0$. This in turn implies that the superposition of $\Phi_1$ and $\Phi_2$ in Eq. (\ref{abc}) in the rest frame, which 
leads to a multi-peak soliton, is not possible in the moving frame. 
 In other words, the multi-peak soliton cannot propagate with
a constant velocity maintaining its shape. On the other hand, for a single-peak
soliton no mixing between $\Phi_1$ and $\Phi_2$ is allowed based on energetic 
considerations as discussed in Sec. \ref{Sec-II} and Sec. \ref{Sec-III}. Hence 
a single-peak soliton can traverse with a constant velocity maintaining its 
shape.

\subsection{Numerical solutions}

We numerically solve the coupled Eqs. (\ref{gps-1})-(\ref{gps-3}) using split-time-step 
Crank-Nicolson method \cite{Wang,Muruganandam} in imaginary and real times. 
The ground state is determined by solving Eqs. (\ref{gps-1})-(\ref{gps-3}) in 
imaginary time. In order to fix both the magnetization and normalization, we use the approach 
discussed in Ref. 
\cite{gautam-1}. 
Accordingly, after each iteration in imaginary time, we
renormalize the component wave functions as
\begin{equation}
\phi_j(x,\tau+\delta\tau) = d_j \phi_j(x,\tau),
\end{equation}
where $d_j$'s satisfy the following relations \cite{gautam-1}
\begin{eqnarray}
d_1d_{-1} &=& d_0^2,\label{norm_const1}\\
d_2d_{-2} &=& d_0^2,\\
d_2d_{-1}^2 &=& d_0^3,\label{norm_const3}
\end{eqnarray}
and
\begin{align}
d_1^8N_2+&d_0^2d_1^6N_1+d_0^4d_1^4N_0+d_0^6d_1^2N_{-1}+d_0^8N_{-2} = 1,\label{norm_const4}\\
2d_1^8N_2&+d_0^2d_1^6N_1-d_0^6d_1^2N_{-1}-2d_0^8N_{-2} = {\cal M}\label{norm_const5}.
\end{align}
Hence, one needs to solve the coupled set of non-linear 
Eqs. (\ref{norm_const4})-(\ref{norm_const5}) to determine $d_0$ and $d_1$, which
can be back substituted in Eqs. (\ref{norm_const1})-(\ref{norm_const3}) to
calculate the remaining normalization constants. The
spatial and time steps employed to generate the numerical results in this paper are
$\delta x = 0.05$ and $\delta t = 0.0000625$, respectively.

\section{Numerical Results}
\label{Sec-IV}

First, we consider an SO-coupled $f=2$ spinor BEC of $10000$  $^{23}$Na atoms
trapped in a harmonic trapping potential with $\omega_x/(2\pi) = 20$ Hz, 
$\omega_y/(2\pi) = \omega_z/(2\pi) =400$ Hz. The oscillator
lengths with these parameters are $l_0 = 4.69~\mu$m
and $l_{yz} = 1.05~\mu$m. This value of $l_0$ has been used in all calculations 
to write  the dimensionless GP equations  (\ref{gps-1})-(\ref{gps-3}).
 The scattering lengths of $^{23}$Na
in total spin $f_{\rm tot} = 0,2$ and $4$ channels are $a_0 = 34.9 a_B, a_2 = 45.8 a_B, 
a_4 = 64.5 a_B$ \cite{Kawaguchi, Ciobanu}, respectively, resulting in 
$c_0 = 242.97, c_1 = 12.06, c_2 = -13.03<20c_1$. Here $a_B$ is the Bohr radius.
The ground state density profile with ${\cal M} = 0$ is shown in Fig. \ref{fig2}(a). 
The multi-peak nature of the solution, obtained as result of the
superposition of two counter-propagating plane waves, is consistent with analytic results
obtained in Sec. \ref{Sec-II}. The solution is dynamically stable and retains its
density and phase profile if  real-time propagation of the dynamics is performed upon small perturbation.
\begin{figure}[!t]
\begin{center}
\includegraphics[trim = 5mm 0mm 3cm 0mm, clip,width=\linewidth,clip]{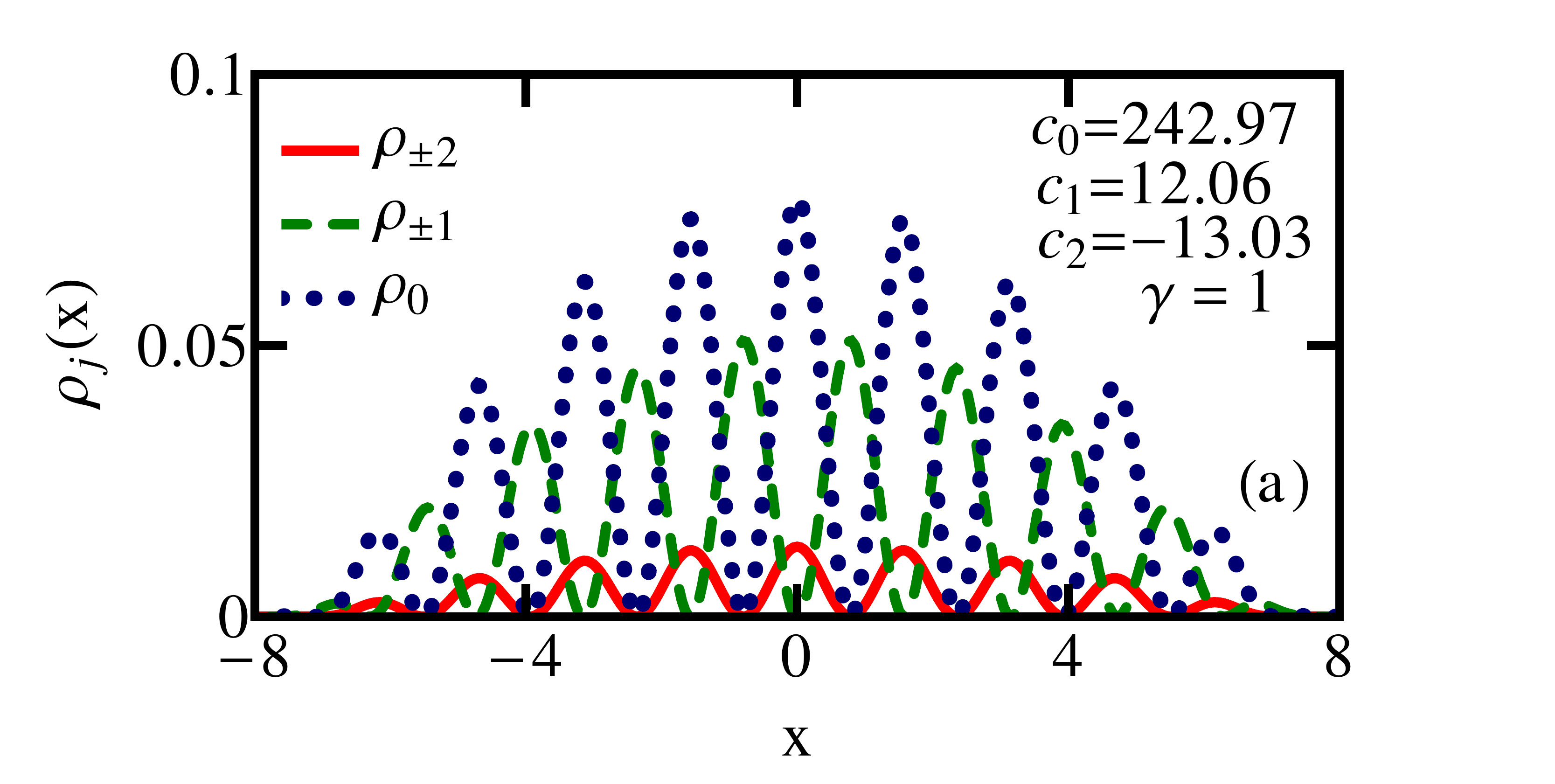}
\includegraphics[trim = 5mm 0mm 3cm 0mm, clip,width=\linewidth,clip]{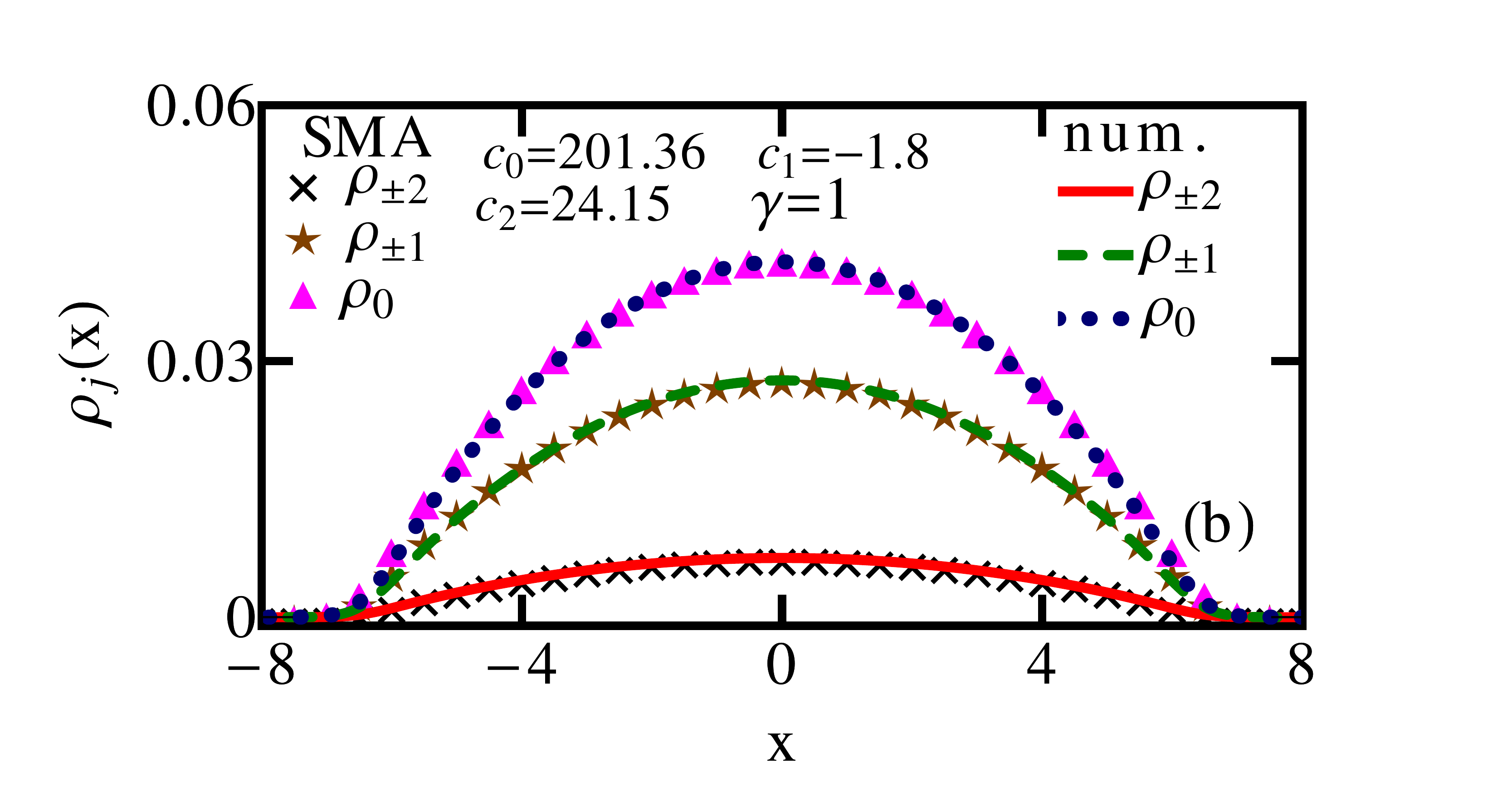}
\caption{(Color online) Ground state density profiles of an SO-coupled trapped 
spinor BEC
 for (a) $c_0 = 242.97, c_1 = 12.06, c_2 = -13.03<20c_1$, and  (b) 
$c_0 = 201.36, c_1 = -1.8, c_2 = 24.15>20c_1$.}
\label{fig2}
\end{center}
\end{figure}

Next we consider the trapped BEC in the $c_2>20c_1$ domain. For this, we consider 
$a_0 = 52.35a_B, a_2= 45.8$, and $a_4= 43a_B$, which results in $c_0 = 201.36, c_1 = -1.8, c_2 = 
24.15>20c_1$. The necessary modification of the scattering lengths can be achieved 
by the optical and magnetic Feshbach resonance techniques \cite{fesh}. 
The trapping potential parameters and number of atoms are the same as those in Fig. \ref{fig2}(a). 
The ground state densities with ${\cal M} = 0$ are shown in 
Fig. \ref{fig2}(b). The component densities obtained by using
SMA, i.e., by solving Eq. (\ref{sma_eq}), are in  good agreement with  the full numerical
solution of Eqs. (\ref{gps-1})-(\ref{gps-3}). The solutions in this case are dynamically stable
plane wave solutions, resulting in a single-peak density distribution for the wave-function
components, consistent with the analytical analysis  in Sec. \ref{Sec-II}.

\begin{figure}[!t]
\begin{center}
\includegraphics[trim = 0mm 0mm 0cm 0mm, clip,width=\linewidth,clip]{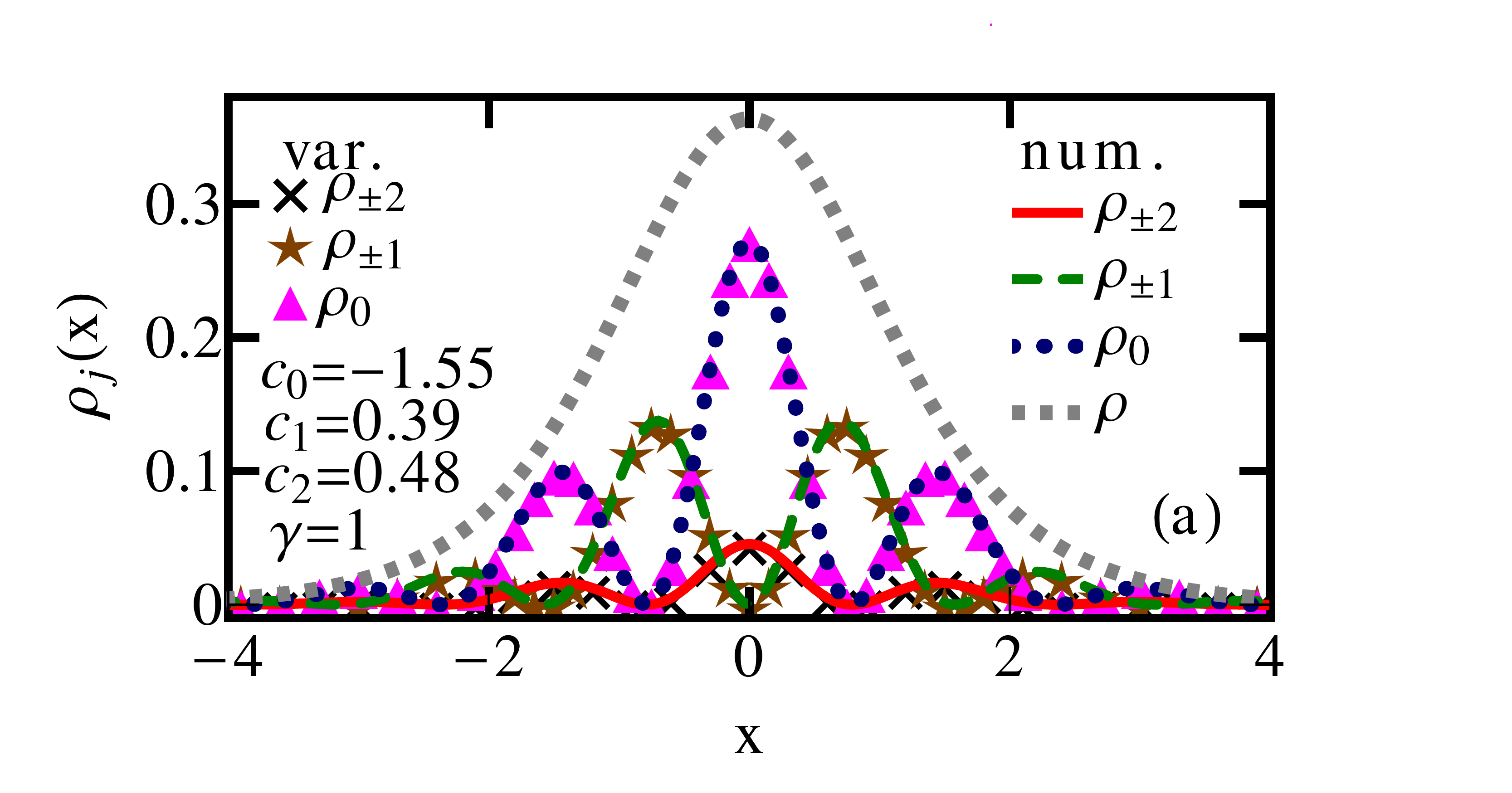}
\includegraphics[trim = 0mm 0mm 0cm 0mm, clip,width=\linewidth,clip]{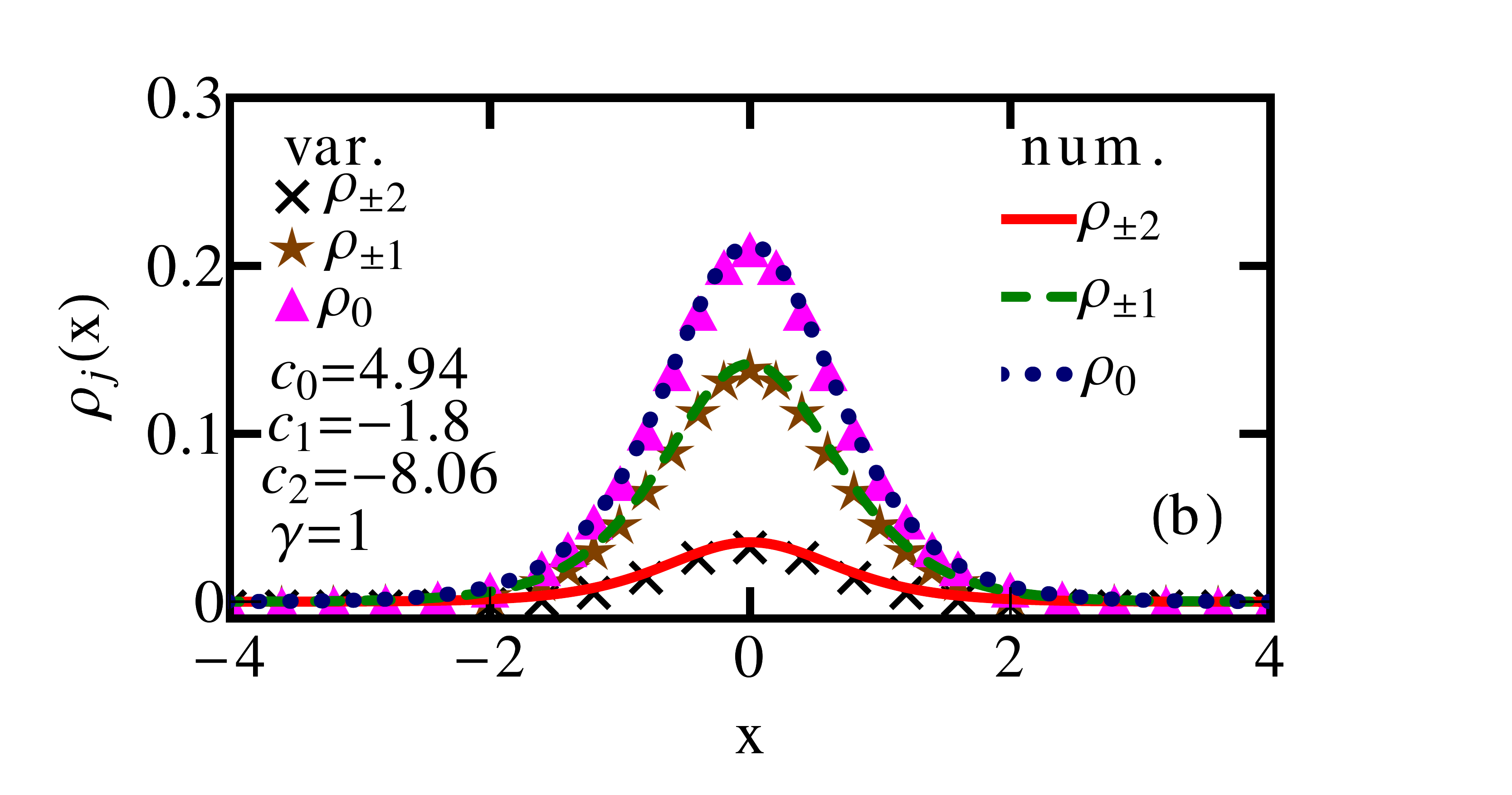}
\includegraphics[trim = 0mm 0mm 0cm 0mm, clip,width=\linewidth,clip]{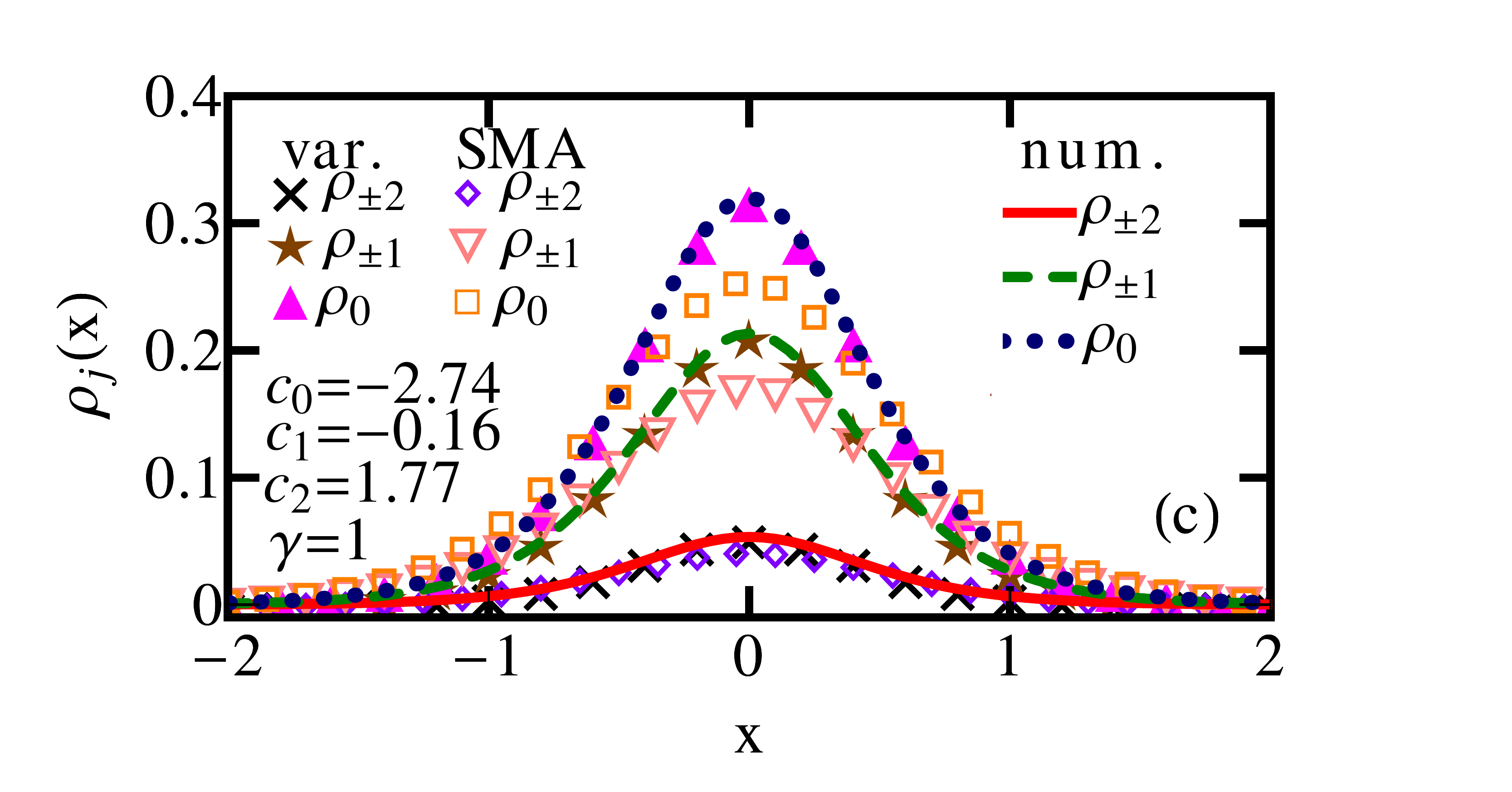}
\caption{(Color online) Numerical (num.)  and variational (var.) densities of an SO-coupled $f=2$  BEC
soliton  
for (a) $c_0 = -1.55, c_1 = 0.39,c_2 = 0.48$,  (b)  
$c_0 = 4.94, c_1 = -1.80,c_2 = -8.06$, and (c) $c_0=-2.74, c_1 = -0.16$,
and $c_2=1.77>20c_1$. In (c) the densities obtained by SMA are also shown. }
\label{fig3} \end{center}
\end{figure}

In order to obtain the bright solitons ($V(x)=0$) in SO-coupled spinor BECs, we consider the two
cases highlighted in Sec. \ref{Sec-IIIA}: (a) $c_2<20c_1$ and   $5c_0+c_2<0$, and (b) $c_2>20c_1$ and
$c_0+4c_1<0$. In case (a), we consider $a_0 = -1.5 a_B$,  
$a_2 = -1.2 a_B$, $a_4 = 0$, and $N = 5000$, which results in $c_0 = -1.55, c_1 = 0.39, 
c_2 = 0.48$. The numerical results for the component densities of the soliton are illustrated 
in Fig. \ref{fig3}(a) together with  the corresponding variational results, defined
by Eqs. (\ref{var_ansatz}) and (\ref{sw1}), with ${\cal M}=0$ and $|\alpha_1| = |\alpha_2| = 1/\sqrt{2}$. The solution in this case has
multi-peak density distribution, is time-reversal {symmetric}, and dynamically stable. 
Despite the modulation in densities of
the individual components, the total density profile ($\rho$) of the bright soliton
as shown in Fig. \ref{fig3}(a) does not have any modulation unlike the bright solitons 
discussed in Ref. \cite{Achilleos}. In this sense { the present} multi-peak solitons are spin-$2$
analogues of the {\em stripe} phase discussed in Ref. \cite{C_Wang,Ho}. 
 { Moreover, in Ref.  \cite{Achilleos} the authors noted component symmetries like Real$(\Psi_\uparrow)=-$
Real$(\Psi_\downarrow)$, and Imag$(\Psi_\uparrow)=$
Imag$(\Psi_\downarrow)$, for the real and imaginary parts, which is not possible in the present model. This is because 
of different SO coupling used in the two studies, e.g.,  $\gamma p_x \Sigma_x$ in this paper and $\gamma p_x \sigma_z$ in Ref.  \cite{Achilleos}. Had we used the SO coupling $\gamma p_x \Sigma_z$ the component symmetries 
of Ref. \cite{Achilleos} can be obtained as noted  in  Table 1 of Ref. \cite{gautam-1}.}

\begin{figure}[!t]
\begin{center}
\includegraphics[trim = 0mm 0mm 0cm 0mm, clip,width=\linewidth,clip]{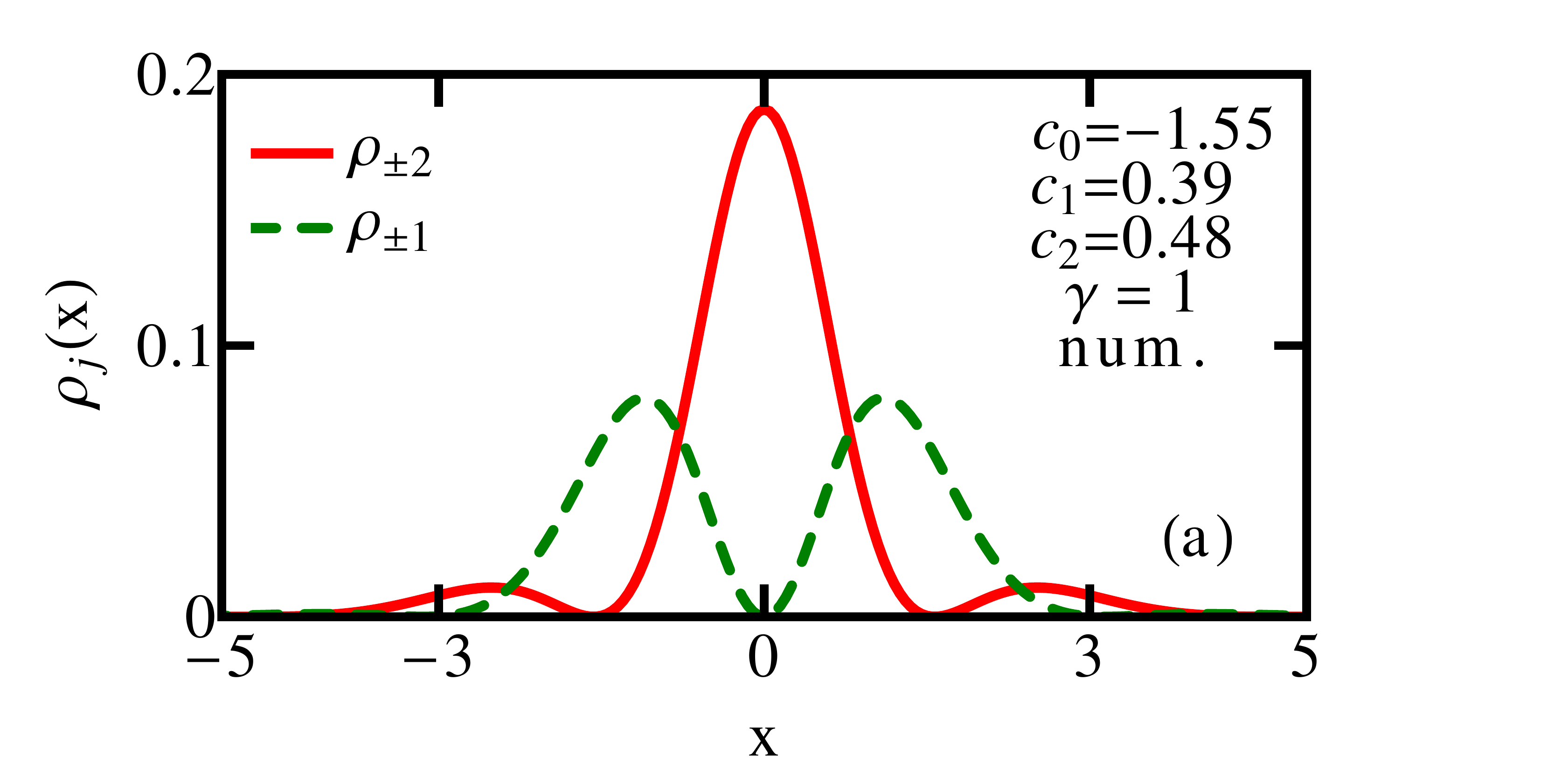}
\includegraphics[trim = 0mm 0mm 0cm 0mm, clip,width=\linewidth,clip]{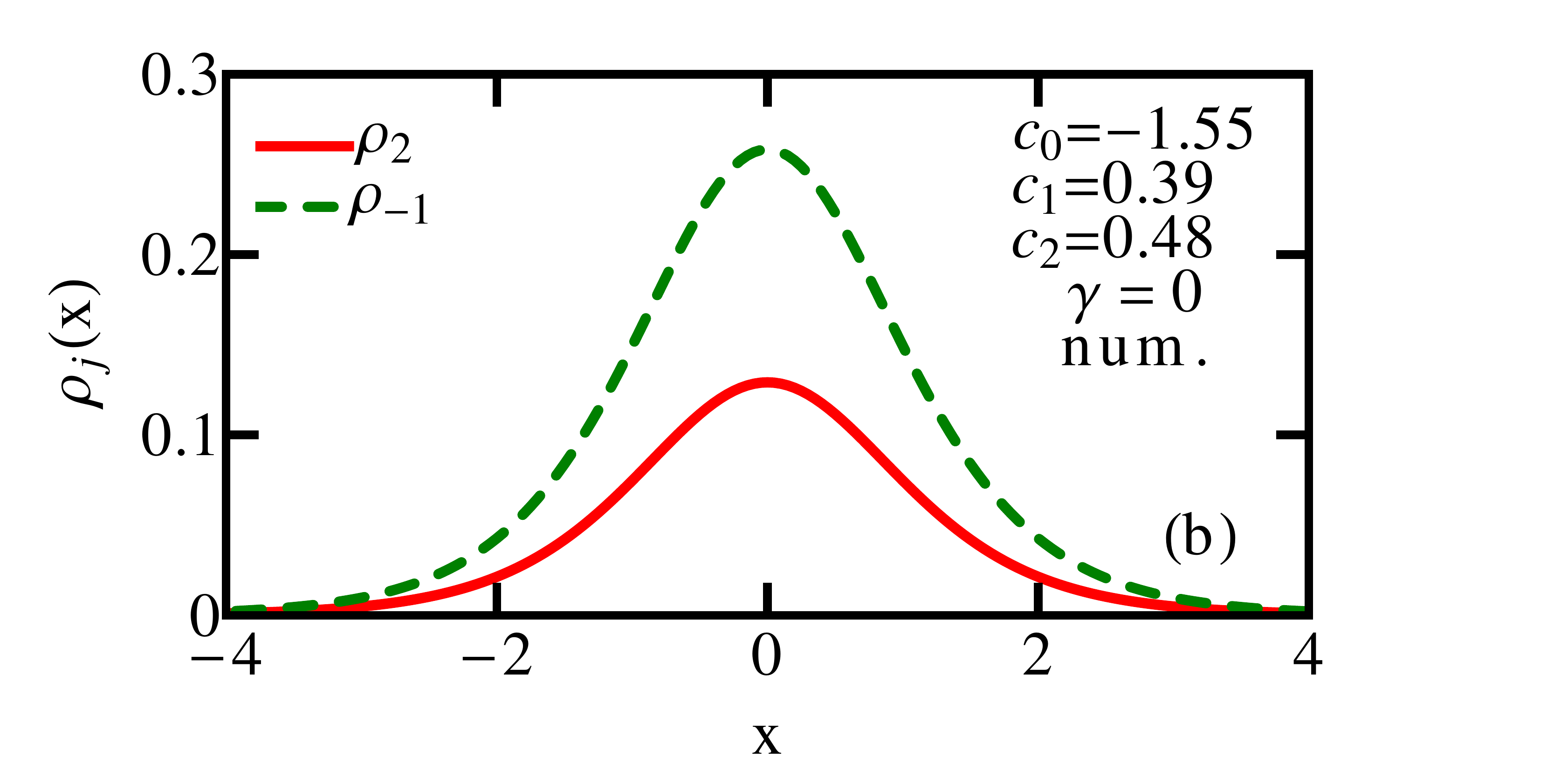}
\caption{(Color online) (a) {Nonzero  numerical  densities} of an SO-coupled $f=2$  BEC
soliton with energy larger than the 
bright soliton shown in Fig. \ref{fig3}(a) .
(b) {Nonzero  numerical  densities} for
an $f=2$ BEC soliton in cyclic phase in the absence of SO coupling. 
The interaction
parameters for both (a) and (b) are $c_0 = -1.55, c_1 = 0.39,c_2 = 0.48$.}
\label{fig4} \end{center}
\end{figure}

In order to obtain the stationary bright soliton with higher energy, we use imaginary time
propagation with $(\Phi_3+\Phi_4)\exp(-x^2/2)/(\sqrt{2\sqrt{\pi}})$ as the initial
guess for the order parameter. The dynamically stable bright soliton  thus obtained 
{ with the parameters 
of Fig. \ref{fig3}(a)} 
is shown in 
Fig. \ref{fig4}(a). The period of density modulation in this case is twice of the
period in Fig. \ref{fig3}(a). Hence, there are two distinct stable multi-peak solitons
in SO-coupled spin-2 condensate as compared to SO-coupled spin-1 condensate which
can have only one type of multi-peak soliton \cite{Liu}.  There also
exists a stable single-peak soliton, not shown here, with order parameter
$\sqrt{3\rho(x)/8}(1,0,-\sqrt{2/3},0,1)$ and energy higher than the aforementioned two multi-peak solitons.
{For the same set of parameters, these three types of solitons $-$ minimum-energy ground and higher-energy excited states $-$  have 
the same total density $\rho(x)$.} 

\begin{figure}
    \includegraphics[trim = 0mm 7mm 0cm 7mm, clip,width=.9\linewidth,clip]{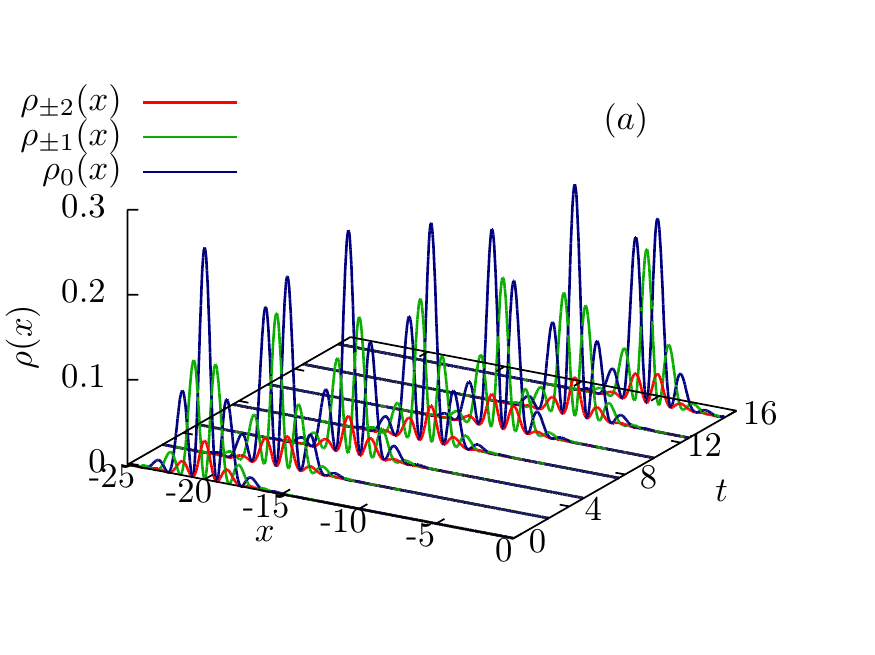}
    \includegraphics[trim = 0mm 7mm 0cm 7mm, clip,width=.9\linewidth,clip]{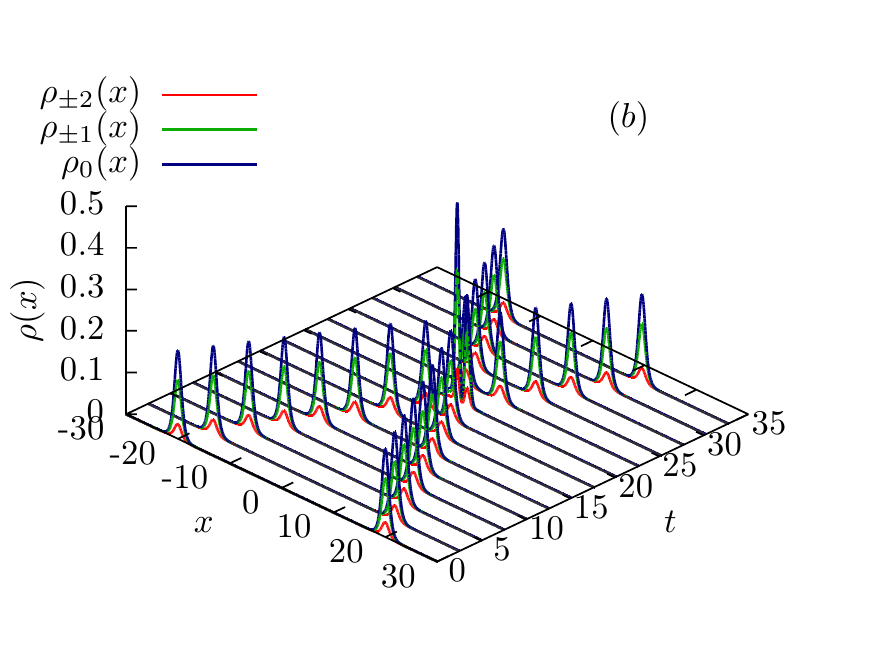}
\caption{(Color online) (a) Dynamics of a soliton of Fig. \ref{fig3}(a)   placed at 
$x=-20$ and set into motion with a constant velocity by multiplying the wave function by a phase  $e^{ix}$.   Here $c_0 = -1.55$, $c_1 =
0.39$, $c_2 = 0.48$, ${\cal M} = 0$, and $v = 1$. (b) Elastic collision 
between the two moving solitons of Fig. \ref{fig3}(b) placed at $x=\pm 20$ and set into motion 
by multiplying by phase factors $e^ {\mp ix}$, respectively. Here $c_0 = 4.94, c_1 = -1.80, c_2 = -8.06$, 
$N = 5000$, ${\cal M} = 0$, and $v = 1$.}
 \label{fig5}
\end{figure}

In the absence of SO coupling the minimum-energy soliton 
corresponds to time-reversal symmetry breaking cyclic phase ($|\mathbf F| = |\Theta| = 0$) 
for $c_1>0$ and $c_2>0,$ { $c_2<20c_1$ \cite{Ciobanu,gautam-1,Kawaguchi}.} This is shown in Fig. \ref{fig4}(b) for the interaction parameters corresponding
to Fig. \ref{fig3}(a) and $\gamma = 0$. The presence of SO coupling results in the
time-reversal symmetric antiferromagnetic phase for which $|\mathbf F| = 0$ and $|\Theta| >0$.
{Next we consider an SO-coupled soliton with   }
$a_0 = 3.4 a_B$,  $a_2 = 4.58 a_B$, $a_4 = -1.0 a_B$, and $N = 5000$, which is
equivalent to $c_0 = 4.94, c_1 = -1.80, c_2 = -8.06$, { $c_2>20c_1$}. The numerical results in this case 
are contrasted  in Fig. \ref{fig3}(b)  with the 
variational 
results, defined by Eqs. (\ref{var_ansatz}) and (\ref{sw2}) with $|\alpha_1| = 1, |\alpha_2| = 0$
or vice versa. %The dynamically stable plane wave 
%solution, which results in a single-peak density distribution in this case, 
{The solitonic solution here 
breaks the 
time-reversal symmetry of the Hamiltonian, as can be seen, for example, from Eqs. 
(\ref{var_ansatz}) and (\ref{eigen_function1}) with $\alpha_2=0.$}
  In the absence of SO coupling
the density distribution remains unchanged in this case for the
bright soliton with zero magnetization, but the component wavefunctions have constant phase instead
of constant phase gradient. The solution thus retains the time-reversal symmetry of the 
Hamiltonian. The SMA can not be applied in this case to calculate the bright solitonic
solution as $c_0>0$. To study the applicability of SMA, we consider $c_0=-2.74, c_1 = -0.16$,
and $c_2=1.77>20c_1$. The bright soliton obtained by  the   numerical solution of Eqs. (\ref{gps-1})-
(\ref{gps-3}), their variational approximation,  viz. Eqs. (\ref{var_ansatz}) and (\ref{sw2}) 
with $|\alpha_1| = 1, |\alpha_2| = 0$, and  SMA,
viz. Eq. (\ref{sma_eq}, 
are shown in Fig. \ref{fig3}(c).

In order to set the soliton into motion with a constant velocity $v$, we multiply
the static solution by $\exp{(ivx)}$ with $v=1$. We find that bright solitons with standing
wave density profile in the $c_2<20c_1$ and $5c_0+c_2<0$ parameter domain, exhibit 
spin-mixing dynamics as is shown in Fig. \ref{fig5}(a) for the soliton with same 
interaction parameters as in Fig. \ref{fig3}(a) and moving with velocity $v=1$ 
in dimensionless units. Thus, this type of soliton does not strictly preserve 
its shape while in motion. On the other hand, moving solitons obtained by 
multiplying the plane wave solutions in the $c_0+4c_1<0$ and $c_2>20c_1$ domain, 
by $\exp(\pm ivx)$ strictly preserve their shape. This is evident from 
Fig. \ref{fig5}(b), where we numerically study the collision between two bright 
solitons, initially located at $x = \pm 20 $, and moving in opposite directions 
with a speed $v = 1$ (in dimensionless units). The solitons collide at $x = 0$ and 
pass through each other without suffering a change in their shapes as is evident 
from Fig. \ref{fig5}(b). The interaction parameters in this case are the same as in 
Fig. \ref{fig3}(b). As discussed in the Sec. \ref{III-B}, these  
moving five-component vector solitons
% stationary solutions of the coupled 
%Eqs. (\ref{gps-1})-(\ref{gps-3}) in a frame moving with  a velocity $v$ and hence 
do not exhibit any spin-mixing dynamics.

\section{Summary}
\label{Sec-V}

We study the generation and propagation of five-component vector solitons  in an SO-coupled spinor  BEC with 
hyperfine spin $f=2$ using a mean-field GP equation  with  three
interaction strengths: $c_0 , c_ 1,$ 
and $c_2$. 
Two types  of solitons $-$ single-peak and multi-peak $-$ emerge in this case for 
  $c_2>20c_1$ and $c_2<20c_1$, respectively. 
In the former case, the solutions  violate  time-reversal symmetry, whereas 
 in the latter  case, the solutions  are time-reversal symmetric. 
The GP equation for this system is  demonstrated to violate  Galelian invariance.  Analyzing the Galelian invariance of this equation, we show  that the single-peak  solitons can move with a constant velocity maintaining constant component densities. On the other hand,  the multi-peak  SO-coupled solitons change the component densities during motion.

%{\em Conclusions:}
%%%%%%%%%%%%%%%%%%%%%%%%%%%%%%%%%%%%%%%%%%%%%%%%%%%%%%%%%%%%%%%%%%%%%%%%%%%%%%%
%%%%%%%%%%%%             Acknowledgements                          %%%%%%%%%%%%
%%%%%%%%%%%%%%%%%%%%%%%%%%%%%%%%%%%%%%%%%%%%%%%%%%%%%%%%%%%%%%%%%%%%%%%%%%%%%%%

\begin{acknowledgements}
This study is financed by the Funda\c c\~ao de Amparo \`a Pesquisa do Estado de 
S\~ao Paulo (Brazil) under contracts  2013/07213-0  and  2012/00451-0 and also by 
the Conselho Nacional de Desenvolvimento Cient\'ifico e Tecnol\'ogico (Brazil) under 
 project 303280/2014-0.
\end{acknowledgements}

\end{document}